\def\ga{\mathrel{\raise.3ex\hbox{$>$\kern-.75em\lower1ex\hbox{$\sim$}}}}
\def\la{\mathrel{\raise.3ex\hbox{$<$\kern-.75em\lower1ex\hbox{$\sim$}}}}
\def\beq{\begin{equation}}
\def\eeq{\end{equation}}

\def\PL{{\it Phys.Lett.} }
\def\PR{{\it Phys.Rev.} }
\def\PRL{{\it Phys.Rev.Lett.} }

\documentclass{ws-procs9x6}

\begin{document}

\title{Searching for Supersymmetric Dark Matter\footnote{Summary of invited talk
at CosPA 2002, 2002 International Symposium on Cosmology and Particle 
Astrophysics, May31 - June 2, 2002, National Taiwan University Taipei, Taiwan.}}

\author{Keith A. Olive}

\address{Theoretical Physics
Institute, School of Physics and Astronomy, \\
University of Minnesota, Minneapolis, MN 55455 USA \\ 
E-mail: olive@umn.edu}


\maketitle

\abstracts{
\vskip -2.5in
\rightline{hep-ph/0208092}
\rightline{UMN--TH--2108/02} 
\rightline{TPI--MINN--02/25}
\rightline{August 2002}
\vskip 1.9in
The supersymmetric extension to the 
Standard Model offers a promising cold dark
matter candidate, the lightest neutralino. I will review the prospects for the detection
of this candidate in both accelerator and direct detection searches.}

\section{Introduction}

Although there are many reasons for
considering supersymmetry as a candidate
extension to the standard model of strong,
weak and electromagnetic interactions\cite{reviews}, one of
the most compelling is its role in
understanding the hierarchy problem\cite{hierarchy} namely, 
why/how is
$m_W \ll M_P$.  One might think naively that it would be
sufficient to set $m_W \ll M_P$ by hand. However, radiative
corrections tend to destroy this hierarchy. For example,
one-loop diagrams generate
\beq
\delta m^2_W = \mathcal {O}\left({\alpha\over\pi}\right)~\Lambda^2 \gg m^2_W
\label{four}
\eeq
where $\Lambda$ is a cut-off representing the appearance of new physics, and
the
inequality in (\ref{four}) applies if $\Lambda\sim 10^3$ TeV, and even 
more so if  $\Lambda \sim m_{GUT} \sim
10^{16}$
GeV or $ \sim M_P \sim 10^{19}$ GeV. If the radiative corrections to a 
physical
quantity
are much larger than its measured values, obtaining the latter requires
strong
cancellations, which in general require fine tuning of the bare input
parameters.
However, the necessary cancellations are natural in supersymmetry, where 
one has
equal
numbers of bosons $B$ and fermions $F$ with equal couplings, so that
(\ref{four})
is replaced by
\beq
\delta m^2_W = \mathcal {O}\left({\alpha\over\pi}\right)~\vert m^2_B - 
m^2_F\vert~.
\label{five}
\eeq
The residual radiative correction is naturally small if
$
\vert m^2_B - m^2_F\vert \la 1~{\rm TeV}^2
$.

In order
to justify the absence of interactions
which can be responsible for extremely rapid proton decay, it
is common in the minimal supersymmetric standard model (MSSM)
to assume the conservation of R-parity.  If R-parity, which
distinguishes between ``normal" matter and the  supersymmetric
partners and can be defined in terms of baryon, lepton and spin
as $R = (-1)^{3B + L + 2S}$, is unbroken, there is at least one 
supersymmetric particle (the lightest supersymmetric particle or LSP)
which must be stable.  Thus, the minimal model contains the fewest number of new
particles and interactions necessary to make a consistent theory.

There are very strong constraints, however, forbidding the existence of stable or
long lived particles which are not color and electrically neutral~\cite{EHNOS}. Strong
and electromagnetically interacting LSPs would become bound with normal matter
forming anomalously heavy isotopes. Indeed, there are very strong upper limits on the
abundances, relative to hydrogen, of nuclear isotopes\cite{isotopes},
$n/n_H \la 10^{-15}~~{\rm to}~~10^{-29}
$
for 1 GeV $\la m \la$ 1 TeV. A strongly interacting stable relic is expected
to have an abundance $n/n_H \la 10^{-10}$
with a higher abundance for charged particles.

There are relatively few supersymmetric candidates which are not colored and
are electrically neutral.  The sneutrino\cite{snu} is one possibility,
but in the MSSM, it has been excluded as a dark matter candidate by
direct\cite{dir} and indirect\cite{indir} searches.  In fact, one can set
an accelerator based limit on the sneutrino mass from neutrino counting, 
$m_{\tilde\nu}\ga$ 44.7 GeV \cite{EFOS}. In this case, the direct relic
searches in
underground low-background experiments require  
$m_{\tilde\nu}\ga$ 20 TeV~\cite{dir}. Another possibility is the
gravitino which is probably the most difficult to exclude. 
I will concentrate on the remaining possibility in the MSSM, namely the
neutralinos.

\section{Parameters}

The most general version of the MSSM, despite its minimality in particles and
interactions contains well over a hundred new parameters. The study of such a model
would be untenable were it not for some (well motivated) assumptions.
These have to do with the parameters associated with supersymmetry breaking.
It is often assumed that, at some unification scale, all of the gaugino masses
receive a common mass, $m_{1/2}$. The gaugino masses at the weak scale are
determined by running a set of renormalization group equations.
Similarly, one often assumes that all scalars receive a common mass, $m_0$,
at the GUT scale. These too are run down to the weak scale. The remaining 
parameters of importance involve the Higgs sector.  There is the Higgs mixing mass
parameter, $\mu$, and since there are two Higgs doublets in the MSSM, there are 
two vacuum expectation values. One combination of these is related to the $Z$ mass,
and therefore is not a free parameter, while the other combination, the ratio of the
two vevs, $\tan \beta$, is free. 

If the supersymmetry breaking Higgs soft masses are also unified at the GUT scale
(and take the common value $m_0$), then $\mu$ and the physical Higgs masses at the
weak scale are determined by electroweak vacuum conditions. This scenario is often
referred to as the constrained MSSM or CMSSM. Once these parameters are set, the
entire spectrum of sparticle masses at the weak scale can be calculated.

\section{Neutralinos}

 There are four neutralinos, each of which is a  
linear combination of the $R=-1$ neutral fermions,\cite{EHNOS}: the wino
$\tilde W^3$, the partner of the
 3rd component of the $SU(2)_L$ gauge boson;
 the bino, $\tilde B$, the partner of the $U(1)_Y$ gauge boson;
 and the two neutral Higgsinos,  $\tilde H_1$ and $\tilde H_2$.
Assuming gaugino mass universality at the  GUT scale, the identity and
mass of the LSP are determined by the gaugino mass $m_{1/2}$, 
$\mu$, and  $\tan \beta$. In general,
neutralinos can  be expressed as a linear combination
\begin{equation}
	\chi = \alpha \tilde B + \beta \tilde W^3 + \gamma \tilde H_1 +
\delta
\tilde H_2
\end{equation}
The solution for the coefficients $\alpha, \beta, \gamma$ and $\delta$
for neutralinos that make up the LSP 
can be found by diagonalizing the mass matrix
\beq
      ({\tilde W}^3, {\tilde B}, {{\tilde H}^0}_1,{{\tilde H}^0}_2 )
  \left( \begin{array}{cccc}
M_2 & 0 & {-g_2 v_1 \over \sqrt{2}} &  {g_2 v_2 \over \sqrt{2}} \\
0 & M_1 & {g_1 v_1 \over \sqrt{2}} & {-g_1 v_2 \over \sqrt{2}} \\
{-g_2 v_1 \over \sqrt{2}} & {g_1 v_1 \over \sqrt{2}} & 0 & -\mu \\
{g_2 v_2 \over \sqrt{2}} & {-g_1 v_2 \over \sqrt{2}} & -\mu & 0 
\end{array} \right) \left( \begin{array}{c} {\tilde W}^3 \\
{\tilde B} \\ {{\tilde H}^0}_1 \\ {{\tilde H}^0}_2 \end{array} \right)
\eeq
where $M_1 (M_2)$ is a soft supersymmetry breaking
 term giving mass to the U(1) (SU(2))  gaugino(s).
  In a unified
 theory $M_1 = M_2$ at the unification scale (at the weak scale, $	M_1 \simeq
{5 \over 3}  {\alpha_1 \over \alpha_2}  M_2	$).   As one can see,  the
coefficients
$\alpha, \beta, \gamma,$ and $\delta$ depend only on
$m_{1/2}$, $\mu$, and $\tan \beta$.

\begin{figure}[th]
	\centering
	\epsfxsize=10cm
\epsfbox{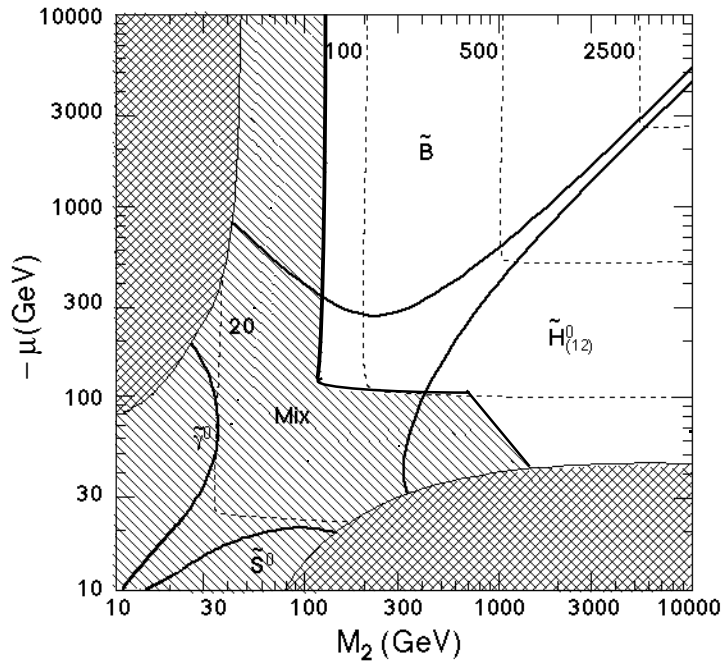}
	\caption{Mass contours and composition of nearly pure LSP states in the MSSM
\protect\cite{osi3}.}
	\label{osi399}
\end{figure}

In Figure \ref{osi399} \cite{osi3}, regions in
the $M_2, \mu$  plane with $\tan\beta = 2$ are shown in which the LSP
is one of several nearly pure states, the photino, $\tilde \gamma$, the
bino,
$\tilde B$, a symmetric combination of the Higgsinos, 
$\tilde{H}_{(12)}$, or the Higgsino, 
$\tilde{S} = \sin \beta {\tilde H}_1 + \cos \beta {\tilde
H}_2$. The dashed lines show the LSP mass contours.
 The cross hatched regions correspond to parameters giving
  a chargino ($\tilde W^{\pm}, \tilde H^{\pm}$) state 
with mass $m_{\tilde \chi} \leq 45 GeV$ and as such are 
excluded by LEP\cite{lep2}.
This constraint has been extended by LEP\cite{LEPsusy} and is shown by
the  light shaded region and corresponds to regions where the chargino
mass is $\la 104$ GeV. The newer limit does not extend deep into the
Higgsino region because of the degeneracy between the chargino and
neutralino.
 Notice that the parameter space is dominated by the  
$\tilde B$ or $\tilde H_{12}$
 pure states and that the photino 
 only occupies a small fraction of the parameter space,
 as does the Higgsino combination $\tilde S$. Both of these
light states are experimentally excluded.

\section{The Relic Density}

The relic abundance of LSP's is 
determined by solving
the Boltzmann
 equation for the LSP number density in an expanding Universe.
 The technique\cite{wso} used is similar to that for computing
 the relic abundance of massive neutrinos\cite{lw}.
The relic density depends on additional parameters in the MSSM beyond $m_{1/2},
\mu$, and $\tan \beta$. These include the sfermion masses, $m_{\tilde f}$ and the
Higgs pseudo-scalar mass, $m_A$\footnote{In general, the relic density depends on
the supersymmetry-breaking tri-linear masses
$A$ (also assumed to be unified at the GUT scale) as well as two phases $\theta_\mu$
and $\theta_A$.}, derived from $m_0$ (and $m_{1/2}$). To determine the relic density
it is necessary to obtain the general annihilation cross-section for neutralinos.  In
much of the parameter space of interest, the LSP is a bino and the
annihilation proceeds mainly through sfermion exchange.
Because of the p-wave suppression associated with Majorana fermions, the s-wave
part of the annihilation cross-section is suppressed by the outgoing fermion masses. 
This means that it is necessary to expand the cross-section to include p-wave
corrections which can be expressed as a term proportional to the temperature if
neutralinos are in equilibrium. Unless the neutralino mass
happens to lie near near a pole, such as $m_\chi \simeq$
$m_Z/2$ or $m_h/2$, in which case there are large contributions to the
annihilation through direct $s$-channel resonance exchange, the dominant
contribution to
the $\tilde{B} \tilde{B}$ annihilation cross section comes from crossed
$t$-channel sfermion exchange.

Annihilations in the early
Universe continue until the annihilation rate
$\Gamma
\simeq \sigma v n_\chi$ drops below the expansion rate given by the Hubble parameter,
$H$.  For particles which annihilate through approximate weak scale interactions, this
occurs when $T \sim m_\chi /20$. Subsequently, the relic density of neutralinos is
fixed relative to the number of relativistic particles. 
As noted above, the number density of neutralinos is tracked by a
Boltzmann-like equation,
\beq
{dn \over dt} = -3{{\dot R} \over R} n - \langle \sigma v \rangle (n^2 -
n_0^2)
\eeq
where $n_0$ is the equilibrium number density of neutralinos.
By defining the quantity $f = n/T^3$, we can rewrite this equation in terms of $x$,
as
\beq
{df \over dx} = m_\chi \left( {1 \over 90} \pi^2 \kappa^2 N \right)^{1/2}
(f^2 - f_0^2)
\eeq
The solution to this equation at late times (small $x$) yields a constant value of
$f$, so that $n \propto T^3$. 
The final relic density expressed as a fraction of the critical energy density 
can be written as\cite{EHNOS}
\beq
\Omega_\chi h^2 \simeq 1.9 \times 10^{-11} \left({T_\chi \over
T_\gamma}\right)^3 N_f^{1/2} \left({{\rm GeV} \over ax_f + {1\over 2} b
x_f^2}\right)
\label{relic}
\eeq 
where $(T_\chi/T_\gamma)^3$ accounts for the subsequent reheating of the
photon temperature with respect to $\chi$, due to the annihilations of
particles with mass $m < x_f m_\chi$ \cite{oss}. The subscript $f$ refers to
values at freeze-out, i.e., when annihilations cease. The coefficients $a$ and $b$
are related to the partial wave expansion of the cross-section, $\sigma v = a + b x +
\dots $. Eq. (\ref{relic} ) results in a very good approximation to the relic density
expect near s-channel annihilation poles,  thresholds and in regions where the LSP is
nearly degenerate with the next lightest supersymmetric particle\cite{gs}.

\section{Phenomenological and Cosmological Constraints}

For the cosmological limits on the relic density I will assume
\begin{equation}   
0.1 \; \le \; \Omega_\chi h^2 \; \le \; 0.3. 
\label{Omegachi}
\end{equation}
The upper limit being a conservative bound based 
only on the lower limit to the
age of the Universe of 12 Gyr. Indeed, most analyses indicate that
$\Omega_{\rm matter} \la 0.4 - 0.5$ and thus it is very likely that
$\Omega_\chi h^2 < 0.2$. One should note that smaller values of
$\Omega_\chi h^2$  are allowed,
since it is quite possible that some of the cold dark matter might not
consist of LSPs.

The calculated relic density is found to have a relevant cosmological 
density over a wide range of susy parameters. For all values of $\tan \beta$, there
is a `bulk' region with relatively low values of $m_{1/2}$ and $m_0$ where
$0.1 < \Omega_\chi h^2 < 0.3$. However there are a number of regions at large values
of $m_{1/2}$ and/or $m_0$ where the relic density is still compatible with the
cosmological constraints. At large values of $m_{1/2}$, the lighter stau, becomes
nearly degenerate with the neutralino and co-annihilations between these particles
must be taken into account\cite{efo}. 
For non-zero values of $A_0$, there are new regions 
for which $\chi-{\tilde t}$  coannihilations are important\cite{stopco}. At large
$\tan \beta$, as one increases
$m_{1/2}$, the pseudo-scalar mass, $m_A$ begins to drop so that there is 
a wide funnel-like region (at all values of $m_0$) such that $2m_\chi \approx m_A$ and
s-channel annihilations become important\cite{funnel,EFGOSi}. Finally, there is a
region at very high $m_0$ where the value of $\mu$ begins to fall and the LSP
becomes more Higgsino-like.  This is known as the `focus point' region\cite{focus}.

\begin{figure}
\centering
	\epsfxsize=10cm
\epsfbox{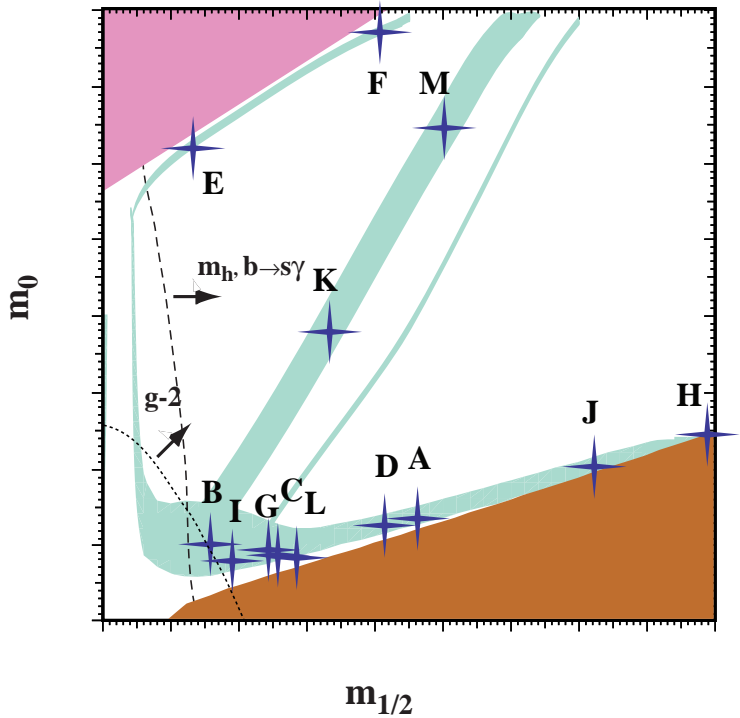}
\caption{Schematic overview of the CMSSM benchmark points proposed 
in~\protect\cite{benchmark}.
 The points are intended to illustrate the range of
available possibilities. The labels correspond to the approximate
positions of the benchmark points in the $(m_{1/2}, m_0)$ plane.
They also span values of $\tan \beta$ from 5  to 50 and include
points with $\mu < 0$. }
\label{fig:Bench}
\end{figure}

As an aid to the assessment of the prospects for detecting sparticles at
different accelerators, benchmark sets of supersymmetric parameters have
often been found useful, since they provide a focus for
concentrated discussion. A set of proposed post-LEP benchmark
scenarios\cite{benchmark} in the CMSSM are illustrated schematically in
Fig.~\ref{fig:Bench}.  Five of the
chosen points are in the
`bulk' region at small $m_{1/2}$ and $m_0$, four are spread along the
coannihilation `tail' at larger $m_{1/2}$ for various values of
$\tan\beta$.  This tail runs along the shaded region in the lower right corner
where the stau is the LSP and is therefore excluded by the constraints against
charged dark matter. 
Two points are in rapid-annihilation `funnels' at large $m_{1/2}$ and $m_0$. 
At large values of $m_0$,  the focus-point region runs along the
boundary where electroweak symmetry no longer occurs (shown in Fig.
\ref{fig:Bench} as the shaded region in the upper left corner). Two points
were chosen in the focus-point region at large $m_0$. The proposed
points range over the allowed values of
$\tan\beta$ between 5 and 50. The light shaded region corresponds to the portion
of parameter space where the relic density $\Omega_\chi h^2$ is
between 0.1 and 0.3.

\begin{figure}[hbtp]
	\centering
		\epsfxsize=10cm
\epsfbox{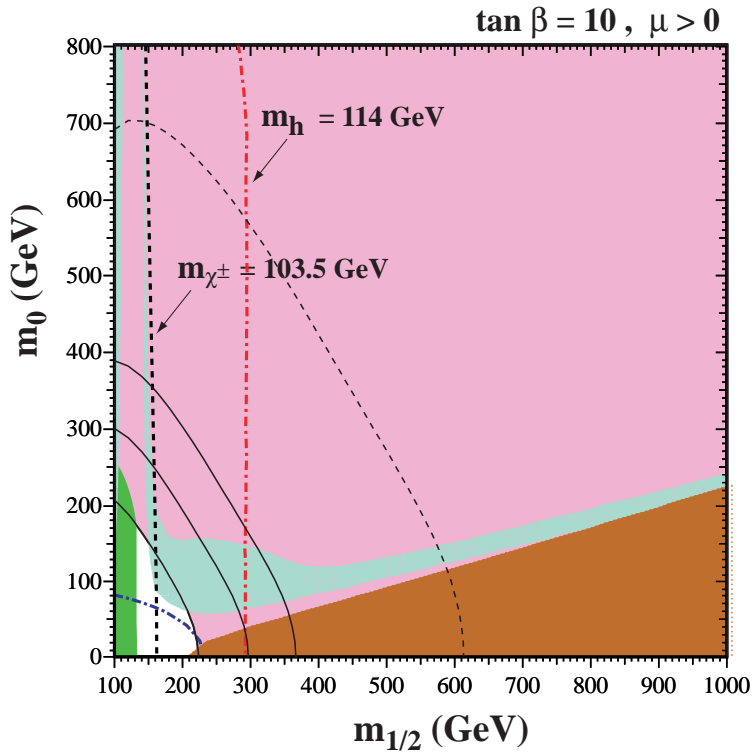}
	\caption{Compilation of phenomenological constraints on the
CMSSM for $\tan \beta = 10,  \mu > 0$,  assuming
$A_0 = 0, m_t = 175$~GeV and $m_b(m_b)^{\overline {MS}}_{SM} =
4.25$~GeV.  The near-vertical lines are the
LEP limits
$m_{\chi^\pm} = 103.5$~GeV (dashed and
black)\protect\cite{LEPsusy}, and
$m_h = 114.1$~GeV (dotted and red)\protect\cite{LEPHiggs}. 
Also, in the lower left corner we show the $m_{\tilde e}
= 99$ GeV contour\protect\cite{LEPSUSYWG_0101}.  In the dark
(brick red) shaded regions, the LSP is the charged
${\tilde \tau}_1$, so this region is excluded. The
light(turquoise) shaded areas are the cosmologically preferred
regions with
\protect\mbox{$0.1\leq \Omega h^2 \leq
0.3$}~\protect\cite{EFGOSi}. The
medium (dark green) shaded regions  are excluded by $b \to s
\gamma$~\protect\cite{bsg}. The shaded (pink) region in the 
upper right delineates the
$2 \, \sigma$ range of $g_\mu - 2$. }
	\label{rd2c}
\end{figure}

The most important phenomenological constraints are also shown schematically in
Figure~\ref{fig:Bench}.  These include the
constraint provided by the LEP lower
limit on the Higgs mass: $m_H > $ 114.1 GeV \cite{LEPHiggs}.
This holds in the Standard Model, for the lightest Higgs boson
$h$ in the general MSSM for
$\tan\beta
\la 8$, and almost always in the CMSSM for all $\tan\beta$.
Since $m_h$ is sensitive to sparticle masses, particularly
$m_{\tilde t}$, via loop corrections,
the Higgs limit also imposes important constraints on the
CMSSM parameters, principally $m_{1/2}$ as seen by the dashed curve in
Fig.~\ref{fig:Bench}. The constraint imposed by
measurements of $b\rightarrow s\gamma$~\cite{bsg} also exclude small values of
$m_{1/2}$. These measurements agree with the Standard Model, and therefore provide
bounds on MSSM particles,  such as the chargino and charged Higgs
masses, in particular. Typically, the $b\rightarrow s\gamma$
constraint is more important for $\mu < 0$, but it is also relevant for
$\mu > 0$,  particularly when $\tan\beta$ is large. 
The BNL E821
experiment reported last year a new measurement of
$a_\mu\equiv {1\over 2} (g_\mu -2)$ which deviated by 2.6 standard
deviations from the best Standard Model prediction available at that
time~\cite{BNL}. The largest contribution to the errors in the comparison
with theory was thought to be the statistical error of the experiment,
which has been significantly reduced just recently \cite{BNL2}. However, it has
recently been realized that the sign of the most important pseudoscalar-meson pole
part of the light-by-light  scattering contribution\cite{lightbylight} to the
Standard Model prediction should be reversed, which reduces the apparent experimental
discrepancy to about 1.6 standard deviations ($\delta a_\mu \times 10^{10}  = 26
\pm 16$). With the new data,  the discrepancy  with theory ranges from 1.6 to 2.6
$\sigma$, i.e., $\delta a_\mu \times 10^{10} = 26
\pm 10$ to $17 \pm 11$\cite{BNL2}.  This constraint excludes very
small values of $m_{1/2}$ and $m_0$. In Fig.~\ref{fig:Bench}, the $g-2$ constraint is
shown schematically by the dotted line. It may also exclude very large values of the
parameters as well as negative values of $\mu$, if the discrepancy holds up.

Following a previous analysis\cite{eos2}, in Figure
\ref{rd2c} the $m_{1/2}-m_0$ parameter space is shown for $\tan
\beta = 10$.   The dark shaded region (in the lower right)
corresponds to the parameters where the LSP is not a neutralino
but rather a
${\tilde \tau}_R$. 
 The cosmologically interesting region at
the left of the figure is due to the appearance
of pole effects.  There, the LSP can annihilate through
s-channel $Z$ and $h$ (the light Higgs) exchange, thereby
allowing a very large value of $m_0$. However, this region is
excluded by phenomenological constraints. Here one can see clearly
the coannihilation tail which extends towards large values of
$m_{1/2}$. In addition to the phenomenological constraints discussed above,
Figure
\ref{rd2c} also shows the current experimental constraints on
the CMSSM parameter space due to the limit $m_{\chi^\pm}
\ga$ 103.5 GeV provided by chargino searches at  LEP
\cite{LEPsusy}. LEP has also provided lower limits on
slepton masses, of which the strongest is $m_{\tilde e}\ga$ 99
GeV \cite{LEPSUSYWG_0101}. This is shown by dot-dashed curve in the
lower left corner of Fig. \ref{rd2c}. 

As one can see, one of the most important phenomenological constraint at this value of
$\tan \beta$ is due to the Higgs mass (shown by the nearly vertical dot-dashed
curve).  The theoretical Higgs masses were evaluated using {\tt
FeynHiggs}\cite{FeynHiggs}, which is estimated to  have a residual uncertainty of a
couple of GeV in $m_h$. The region excluded
by the  $b\rightarrow s\gamma$ constraint is the dark shaded (green)
region to the left of the plot. 

As many authors have pointed out\cite{susygmu}, a discrepancy between
theory and the BNL experiment could well be explained by supersymmetry. As
seen in Fig.~\ref{rd2c}, this is particularly easy if $\mu > 0$.
The
medium (pink) shaded region in the figure corresponds to the overall allowed region
by the new experimental result:
$-5 < \delta a_\mu \times 10^{10} <  46 $. The two solid lines within the shaded
region corresponds to the central values
$\delta a_\mu \times 10^{10} = 17 $ and 26 respectively.  The optimistic $2 \sigma$
lower bound of $\delta a_\mu \times 10^{10} = 6 $ is shown as a dashed curve.

\begin{figure}[hbtp]
	\centering 
		\epsfxsize=10cm
\epsfbox{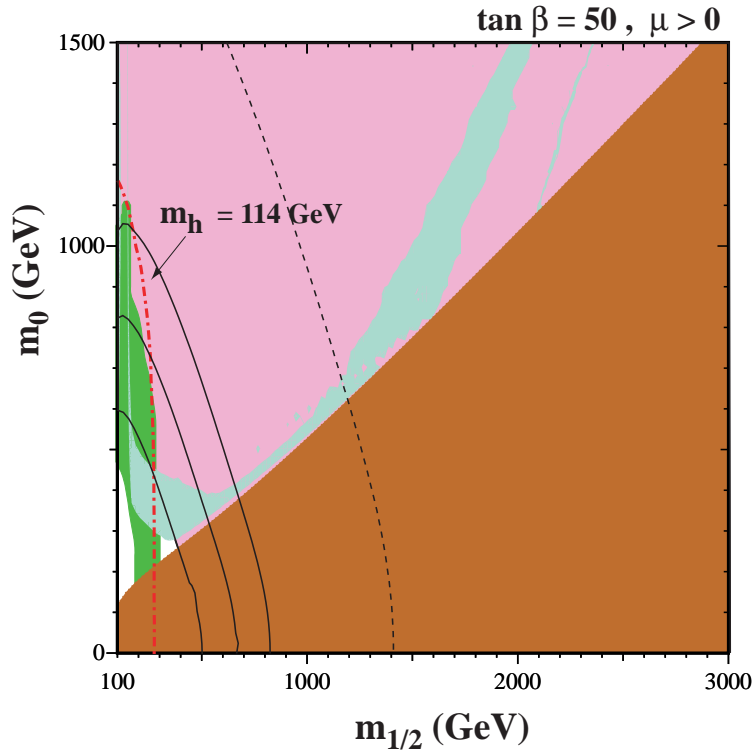}
	\caption{As in Fig. \protect\ref{rd2c} for $\tan \beta = 50$. }
	\label{rd2c50}
\end{figure}

As discussed above, another
mechanism for extending the allowed CMSSM region to large
$m_\chi$ is rapid annihilation via a direct-channel pole when $m_\chi
\sim {1\over 2} m_{A}$~\cite{funnel,EFGOSi}. This may yield a
`funnel' extending to large $m_{1/2}$ and $m_0$ at large $\tan\beta$, as
seen in Fig.~\ref{rd2c50}.

In principle the true 
input parameters in the CMSSM are: $\mu, m_1, m_2,$ and $ B$, where $m_1$ and $m_2$
are the Higgs soft masses (in the CMSSM $m_1 = m_2 = m_0$ and $B$ is the
susy breaking bilinear mass term). In this case, the electroweak symmetry breaking
conditions lead to	a prediction of $M_Z, \tan \beta$ ,and  $m_A$.  Since we are 
not really interested in predicting $M_Z$, it is more useful to assume  
instead the following CMSSM	input parameters: $M_Z, m_1, m_2,$ and $\tan \beta$
again with	$m_1 = m_2 = m_0$. In this case, one predicts $\mu, B$, and $m_A$.
However, one can generalize the CMSSM case to include non-universal Higgs masses
(NUHM), in which case the 			
input parameters become:$ M_Z, \mu, m_A,$ and $\tan \beta$ and one	predicts $m_1,
m_2$, and $B$. 

\begin{figure}[hbtp]
	\centering 
		\epsfxsize=10cm
\epsfbox{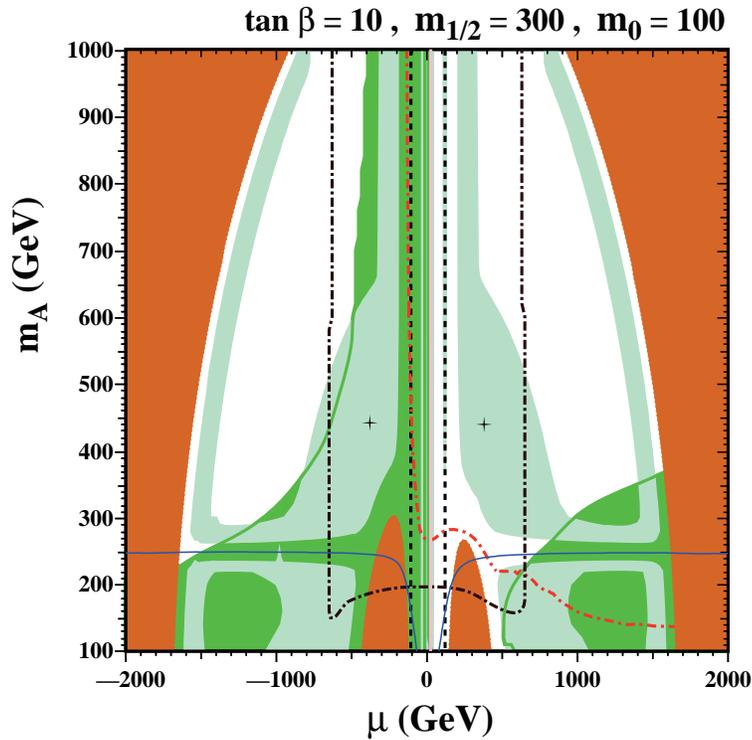}
	\caption{Compilations of phenomenological constraints on the MSSM with NUHM 
in the $(\mu, m_A)$ plane for $\tan \beta = 10$ and $m_0 = 100$~GeV, 
$m_{1/2} = 300$~GeV, assuming $A_0 = 0$, $m_t = 175$~GeV and 
$m_b(m_b)^{\overline {MS}}_{SM} = 4.25$~GeV.
The  shading is as described in Fig.~\protect\ref{rd2c}.
The (blue) solid line is the contour $m_\chi = m_A/2$, near which
rapid direct-channel annihilation suppresses the relic density. 
The dark (black) dot-dashed line indicates when one or another Higgs 
mass-squared becomes negative at the GUT scale: only lower $|\mu|$ and 
larger $m_A$ values are allowed. The crosses denote the values of 
$\mu$ and $m_A$ found in the CMSSM.}
	\label{muma}
\end{figure}

The NUHM parameter space was recently analyzed\cite{eos3} and a sample of the results
found is shown in Fig. \ref{muma}. While much of the cosmologically preferred area
with $\mu < 0$ is excluded, there is a significant enhancement in the allowed 
parameter space for $\mu > 0$. 

\subsection{Detection}

Because the LSP as dark matter is present locally, there are many
avenues for pursuing dark matter detection. Direct detection techniques
rely on an ample neutralino-nucleon scattering cross-section.
The effective four-fermion lagrangian can be written as
\begin{eqnarray}
\mathcal{L}  & =  &\bar{\chi} \gamma^\mu \gamma^5 \chi \bar{q_{i}} 
\gamma_{\mu} (\alpha_{1i} + \alpha_{2i} \gamma^{5}) q_{i}  \nonumber \\
& + & \alpha_{3i} \bar{\chi} \chi \bar{q_{i}} q_{i} + 
\alpha_{4i} \bar{\chi} \gamma^{5} \chi \bar{q_{i}} \gamma^{5} q_{i} \nonumber \\
& + &\alpha_{5i} \bar{\chi} \chi \bar{q_{i}} \gamma^{5} q_{i} +
\alpha_{6i} \bar{\chi} \gamma^{5} \chi \bar{q_{i}} q_{i} 
\end{eqnarray}
However, the terms involving $\alpha_{1i}, \alpha_{4i}, \alpha_{5i}
$, and
$\alpha_{6i}$  lead to velocity dependent  elastic cross sections.
The remaining terms are: the spin dependent coefficient,
$\alpha_{2i} $  and the scalar coefficient $\alpha_{3i} $.
Contributions to $\alpha_{2i} $ are predominantly through light squark exchange. 	
This is the dominant channel for binos.
Scattering also occurs through Z exchange but this channel 
requires a strong Higgsino component.
Contributions to $\alpha_{3i} $  are also dominated by
light squark exchange 	
but Higgs exchange
is non-negligible in most cases. 

\begin{figure}[hbtp]
	\centering 
		\epsfxsize=10cm
\epsfbox{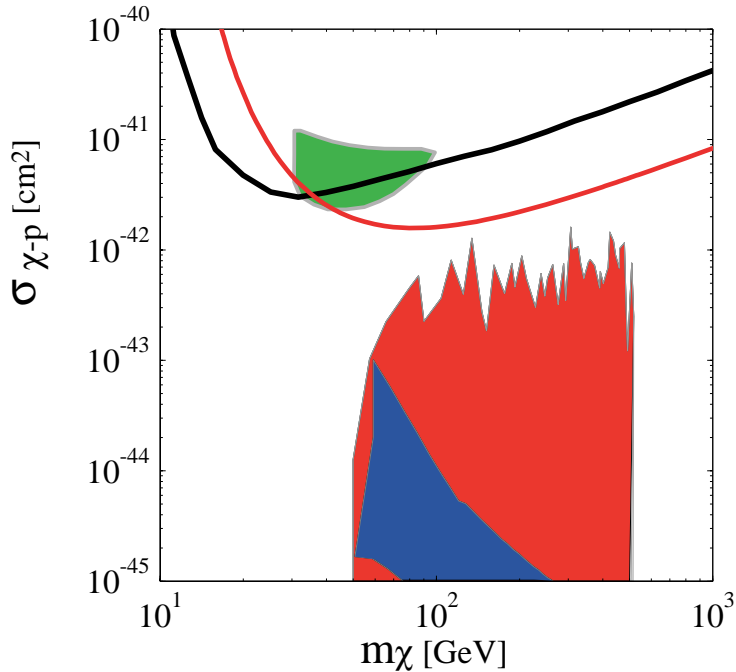}
	\caption{Limits from the CDMS \protect\cite{cdms} and Edelweiss \protect\cite{edel}
experiments on the neutralino-proton elastic scattering cross section as a function
of the neutralino mass. The Edelweiss limit is stronger at higher $m_\chi$. These
results nearly exclude the shaded region observed by DAMA \protect\cite{dama}. The
theoretical predictions lie at lower values of the cross section.}
	\label{cdms}
\end{figure}

The results from a CMSSM and MSSM analysis\cite{EFlO1,EFlO2} for $\tan \beta = 3$ and
10 are compared with the most recent CDMS\cite{cdms} and Edelweiss\cite{edel}
bounds in Fig.~\ref{cdms}. These results have nearly entirely excluded the
region purported by the DAMA\cite{dama} experiment. The CMSSM prediction
\cite{EFlO1} is shown  by the dark shaded region, while the NUHM case \cite{EFlO2} is
shown by the larger lighter shaded region.

\begin{figure}
\centering
		\epsfxsize=10cm
\epsfbox{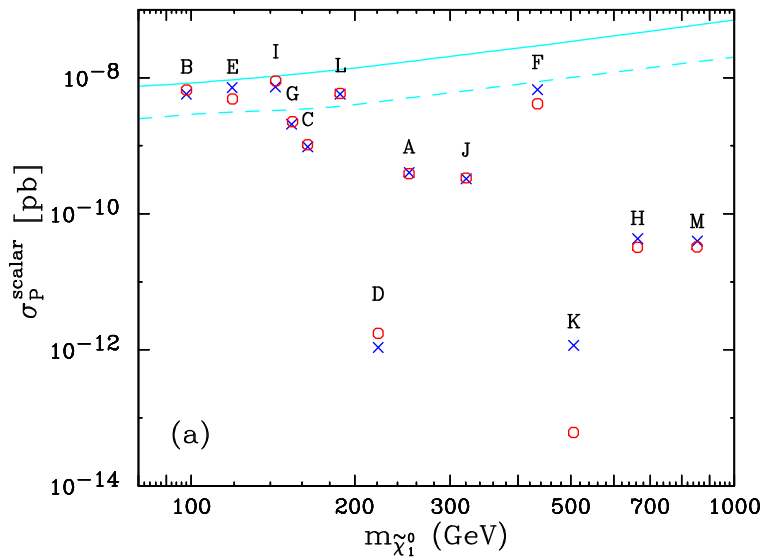}
\caption{ Elastic spin-independent scattering  
of supersymmetric relics on protons  calculated in 
benchmark scenarios\protect\cite{EFFMO}, compared with the 
projected sensitivities for CDMS
II~\protect\cite{Schnee:1998gf} and CRESST\protect\cite{Bravin:1999fc}
(solid) and GENIUS\protect\cite{GENIUS} (dashed).
The predictions of our code (blue
crosses) and {\tt Neutdriver}\protect\cite{neut} (red circles) for
neutralino-nucleon scattering are compared.
The labels A, B, ...,L correspond to the benchmark points as shown in 
Fig.~\protect\ref{fig:Bench}.}
\label{fig:DM}
\end{figure}

I conclude by showing the
prospects for direct detection for the benchmark points discussed
above\cite{EFFMO}. Fig.~\ref{fig:DM} shows rates for the elastic
spin-independent scattering of supersymmetric relics,
including the projected sensitivities for CDMS
II~\cite{Schnee:1998gf} and CRESST\cite{Bravin:1999fc} (solid) and
GENIUS\cite{GENIUS} (dashed).
Also shown are the cross sections 
calculated in the proposed benchmark scenarios discussed in the previous
section, which are considerably below the DAMA\cite{dama} range
($10^{-5} - 10^{-6}$~pb).
Indirect searches for supersymmetric dark matter via the products of
annihilations in the galactic halo or inside the Sun also have prospects
in some of the benchmark scenarios\cite{EFFMO}.

\section*{Acknowledgments}
I would like to thank J. Ellis, T. Falk, A. Ferstl, G. Ganis, Y. Santoso, and  M.
Srednicki for enjoyable collaborations from which this work is culled.
 This work was supported in part by 
DOE grant DE-FG02-94ER40823 at Minnesota.


\begin{thebibliography}{0}

\bibitem{reviews} J.~Wess and J.~Bagger,
{\em Supersymmetry and Supergravity},
(Princeton University Press, Princeton NJ, 1992); \\
G.G.~Ross, {\em Grand Unified Theories},
(Addison-Wesley, Redwood City CA, 1985); \\
S. Martin, arXiv:hep-ph/9709356; \\
J. Ellis, arXiv:hep-ph/9812235; \\
K.A. Olive, arXiv:hep-ph/9911307 .

\bibitem{hierarchy}
L.~Maiani, {\it Proceedings of the 1979 Gif-sur-Yvette Summer School On
Particle
Physics}, 1; \\
G.~'t Hooft, in {\it Recent Developments in Gauge Theories, Proceedings
of the Nato Advanced Study
Institute, Cargese, 1979}, eds. G.~'t Hooft {\it et al.}, (Plenum Press,
NY, 1980); \\ E.~Witten, {\it 
Phys.\ Lett.\ }  {\bf B105}, 267 (1981).


\bibitem{EHNOS}
J. Ellis, J.S. Hagelin, D.V. Nanopoulos, K.A. Olive
and M. Srednicki, {\it Nucl. Phys.}  {\bf B238}, 453 (1984); \\ see also
H. Goldberg, {\it Phys. Rev. Lett.} {\bf 50}, 1419 (1983).

\bibitem{isotopes} J. Rich, M. Spiro and J. Lloyd-Owen, {\it Phys.Rep.}
{\bf 151}, 239 (1987); \\
P.F. Smith, {\it Contemp.Phys.} {\bf 29}, 159 (1998); \\
T.K. Hemmick et al., \PR {\bf D41}, 2074 (1990).

\bibitem{snu} L.E. Ibanez, {\it Phys. Lett.} {\bf 137B}, 160 (1984); \\
	J. Hagelin, G.L. Kane, and S. Raby, 
{\it Nucl., Phys.} {\bf B241}, 638 (1984); \\
T. Falk, K.A. Olive, and M. Srednicki, 
{\it Phys. Lett.} {\bf B339}, 248 (1994).

\bibitem{dir}S. Ahlen, et. al., {\it Phys. Lett.} {\bf B195}, 603 (1987);
\\
	D.D. Caldwell, et. al., {\it Phys. Rev. Lett.} {\bf 61}, 510 (1988); \\
M. Beck et al., {\it Phys. Lett.} {\bf B336} 141 (1994).

\bibitem{indir} see e.g.	K.A. Olive and M. Srednicki,
 {\it Phys. Lett.} {\bf 205B}, 553 (1988).


\bibitem{EFOS} The LEP Collaborations, the LEP Electroweak Working Group, and the SLD
Heavy Flavour and Electroweak Groups, CERN-EP-2000-016.


\bibitem{osi3} K.A. Olive and M. Srednicki, {\it Phys. Lett.} {\bf B230},
 78 (1989);
{\it Nucl. Phys.} {\bf  B355}, 208 (1991).




\bibitem{lep2} ALEPH collaboration, D. Decamp et al., 
Phys. Rep. {\bf 216}, 253 (1992); \\
L3 collaboration, M. Acciarri et al., 
Phys. Lett. {\bf B350}, 109 (1995); \\
OPAL collaboration, G. Alexander et al., 
Phys. Lett. {\bf B377}, 273 (1996).


\bibitem{LEPsusy}
Joint LEP~2 Supersymmetry Working Group,
{\it Combined LEP Chargino Results, up to 208 GeV}, \\
{\tt http://lepsusy.web.cern.ch/lepsusy/www/inos{\_}moriond01/%
charginos{\_}pub.html}.


\bibitem{wso}R. Watkins, M. Srednicki and K.A. Olive, 
{\it Nucl. Phys.} {\bf  B310},
 693 (1988).

\bibitem{lw}P. Hut, {\it Phys. Lett.} {\bf  69B}, 85 (1977); \\
	B.W. Lee and S. Weinberg, {\it Phys. Rev. Lett.} {\bf  39}, 165 (1977).


\bibitem{oss} G.~Steigman, K.~A.~Olive and D.~N.~Schramm,
{\it Phys.\ Rev.\ Lett.\ } {\bf 43}, 239 (1979); \\
K.~A.~Olive, D.~N.~Schramm and G.~Steigman,
{\it Nucl.\ Phys.\ B} {\bf 180}, 497 (1981).

\bibitem{gs}K. Griest and D. Seckel, \PR {\bf D43}, 3191 (1991).

\bibitem{efo} J. Ellis, T. Falk, and K. Olive, \PL {\bf B444},
367 (1998); \\
J. Ellis, T. Falk, K. Olive, and M. Srednicki, {\it Astr. Part. Phys.} (in
 {\bf 13}, 181 (2000)
[Erratum-ibid.\  {\bf 15}, 413 (2000)];
M.~E.~G\'omez, G.~Lazarides and C.~Pallis,
Phys.\ Rev.\ D {\bf 61} (2000) 123512 
and
Phys.\ Lett.\ B {\bf 487} (2000) 313; \\
R.~Arnowitt, B.~Dutta and Y.~Santoso, 
Nucl. Phys. B {\bf 606} (2001) 59.

\bibitem{stopco}
C.~Boehm, A.~Djouadi and M.~Drees,
{\it Phys.\ Rev.\ D} {\bf 62}, 035012  (2000); \\
J. Ellis, K.A. Olive and Y. Santoso,
arXiv:hep-ph/0112113.



\bibitem{funnel}
M.~Drees and M.~M.~Nojiri,
{\it Phys.\ Rev.\ D} {\bf 47}, 376 (1993); \\
H.~Baer and M.~Brhlik,
{\it Phys.\ Rev.\ D} {\bf 53}, 597  (1996) ;
and {\it Phys.\ Rev.\ D} {\bf 57} 567 (1998); \\
H.~Baer, M.~Brhlik, M.~A.~Diaz, J.~Ferrandis, P.~Mercadante, P.~Quintana
and X.~Tata,
{\it Phys.\ Rev.\ D} {\bf 63}, 015007  (2001); \\
A.~B.~Lahanas, D.~V.~Nanopoulos and V.~C.~Spanos,
{\it Mod. Phys. Lett. A} {\bf 16} 1229 (2001).

\bibitem{EFGOSi}
J.~R.~Ellis, T.~Falk, G.~Ganis, K.~A.~Olive and M.~Srednicki,
{\it Phys.\ Lett.\ B} {\bf 510}, 236 (2001).


\bibitem{focus}
J.~L.~Feng, K.~T.~Matchev and T.~Moroi,
{\it Phys.\ Rev.\ Lett.\ } {\bf 84}, 2322 (2000); \\
J.~L.~Feng, K.~T.~Matchev and T.~Moroi,
{\it Phys.\ Rev.\ D}{\bf 61}, 075005 (2000); \\
J.~L.~Feng, K.~T.~Matchev and F.~Wilczek,
{\it Phys.\ Lett.\ B } {\bf 482}, 388 (2000).

\bibitem{benchmark}
M.~Battaglia et al., {\it Eur. Phys. J. C} {\bf 22} 535 (2001).

\bibitem{LEPHiggs}
LEP Higgs Working Group for Higgs boson searches, OPAL Collaboration,
ALEPH Collaboration, DELPHI Collaboration and L3
Collaboration,
{\it Search for the Standard Model Higgs Boson at LEP},
ALEPH-2001-066, DELPHI-2001-113, CERN-L3-NOTE-2699, OPAL-PN-479,
LHWG-NOTE-2001-03, CERN-EP/2001-055, arXiv:hep-ex/0107029;
{\it Searches for the neutral Higgs bosons of the MSSM: Preliminary
combined results using LEP data collected at energies up to 209 GeV},
LHWG-NOTE-2001-04, ALEPH-2001-057, DELPHI-2001-114, L3-NOTE-2700,
OPAL-TN-699, arXiv:hep-ex/0107030.


\bibitem{bsg}
M.S. Alam et al., [CLEO Collaboration], {\it Phys.\ Rev.\ Lett.\ } {\bf
74}, 2885  (1995); \\ as updated in
S.~Ahmed et al., {CLEO CONF 99-10};
BELLE Collaboration, BELLE-CONF-0003, contribution to the 30th 
International conference on High-Energy Physics, Osaka, 2000; \\
See also
K.~Abe {\it et al.},  [Belle Collaboration],
[arXiv:hep-ex/0107065]; \\
L.~Lista  [BaBar Collaboration],
[arXiv:hep-ex/0110010];\\
C. Degrassi, P. Gambino and G.~F. Giudice,
{\it JHEP} {\bf 0012}, 009  (2000); \\
M.~Carena, D.~Garcia, U.~Nierste and C.~E.~Wagner,
{\it Phys. Lett. B} {\bf 499}, 141  (2001); \\
P.~Gambino and M.~Misiak,
Nucl.\ Phys.\ B {\bf 611} (2001) 338; \\
D.~A.~Demir and K.~A.~Olive,
Phys.\ Rev.\ D {\bf 65}, 034007 (2002).












\bibitem{BNL}
H.~N.~Brown {\it et al.}  [Muon g-2 Collaboration],
{\it Phys.\ Rev.\ Lett.\ } {\bf 86}, 2227 (2001).

\bibitem{BNL2} G.W. Bennet {\it et al.}  [Muon g-2 Collaboration],
arXiv:hep-ex/0208001.


\bibitem{lightbylight}
M.~Knecht and A.~Nyffeler,
Phys.\ Rev.\ D {\bf 65}, 073034 (2002); \\
M.~Knecht, A.~Nyffeler, M.~Perrottet and E.~De Rafael,
Phys.\ Rev.\ Lett.\  {\bf 88}, 071802 (2002); \\
M.~Hayakawa and T.~Kinoshita,
arXiv:hep-ph/0112102; \\
I.~Blokland, A.~Czarnecki and K.~Melnikov,
Phys.\ Rev.\ Lett.\  {\bf 88}, 071803 (2002); \\
J.~Bijnens, E.~Pallante and J.~Prades,
Nucl.\ Phys.\ B {\bf 626}, 410 (2002).



\bibitem{eos2} J.~R.~Ellis, K.~A.~Olive and Y.~Santoso,
New Jour.\ Phys.\  {\bf 4}, 32 (2002).




\bibitem{LEPSUSYWG_0101}
Joint LEP~2 Supersymmetry Working Group, 
{\it Combined LEP
Selectron/Smuon/Stau Results, 183-208 GeV}, \\
{\tt http://alephwww.cern.ch/\~{}ganis/SUSYWG/SLEP/sleptons{\_}2k01.html}.


\bibitem{FeynHiggs}
S.~Heinemeyer, W.~Hollik and G.~Weiglein,
{\it Comput.\ Phys.\ Commun.\ } {\bf 124}, 76 (2000); \\
S.~Heinemeyer, W.~Hollik and G.~Weiglein,
{\it Eur.\ Phys.\ J.\ C} {\bf 9}, 343  (1999).



\bibitem{susygmu}
L.~L.~Everett, G.~L.~Kane, S.~Rigolin and L.~Wang,
{\it Phys.\ Rev.\ Lett.\ } {\bf 86}, 3484 (2001); \\
J.~L.~Feng and K.~T.~Matchev,
{\it Phys.\ Rev.\ Lett.\ } {\bf 86}, 3480 (2001); \\
E.~A.~Baltz and P.~Gondolo,   
{\it Phys.\ Rev.\ Lett.\ } {\bf 86}, 5004 (2001); \\
U.~Chattopadhyay and P.~Nath,
{\it Phys.\ Rev.\ Lett.\ } {\bf 86}, 5854 (2001); \\
S.~Komine, T.~Moroi and M.~Yamaguchi,
{\it Phys.\ Lett.\ B} {\bf 506}, 93 (2001); \\
J.~Ellis, D.~V.~Nanopoulos and K.~A.~Olive,
{\it Phys.\ Lett.\ B} {\bf 508}, 65 (2001); \\
R.~Arnowitt, B.~Dutta, B.~Hu and Y.~Santoso,
{\it Phys.\ Lett.\ B} {\bf 505}, 177 (2001); \\
S.~P.~Martin and J.~D.~Wells,
{\it Phys.\ Rev.\ D} {\bf 64}, 035003 (2001); \\
H.~Baer, C.~Balazs, J.~Ferrandis and X.~Tata,
{\it Phys.\ Rev.\ D} {\bf 64}, 035004 (2001).


\bibitem{eos3} J.~Ellis, K.~Olive and Y.~Santoso,
Phys.\ Lett.\ B {\bf 539}, 107 (2002).

\bibitem{EFlO1}
J.~R.~Ellis, A.~Ferstl and K.~A.~Olive,
Phys.\ Lett.\ B {\bf 481}, 304 (2000);
see also:
Phys.\ Lett.\ B {\bf 532}, 318 (2002).



\bibitem{EFlO2}
J.~R.~Ellis, A.~Ferstl and K.~A.~Olive,
Phys.\ Rev.\ D {\bf 63}, 065016 (2001).

\bibitem{cdms}
D.~Abrams {\it et al.}  [CDMS Collaboration],
arXiv:astro-ph/0203500.

\bibitem{edel}
R.~Jakob,
arXiv:hep-ph/0206271.



\bibitem{dama}
DAMA Collaboration, R.~Bernabei {\it et al.},
{\it Phys.\ Lett.\ B } {\bf 436} (1998) 379.



\bibitem{EFFMO}
J.~Ellis, J.~L.~Feng, A.~Ferstl, K.~T.~Matchev and K.~A.~Olive, 
Eur.\ Phys.\ J.\ C {\bf 24}, 311 (2002)
[arXiv:astro-ph/0110225].




\bibitem{Schnee:1998gf}
CDMS Collaboration, R.~W.~Schnee {\it et al.},
{\it Phys.\ Rept.\ } {\bf 307}, 283 (1998).


\bibitem{Bravin:1999fc}
CRESST Collaboration, M.~Bravin {\it et al.},
{\it Astropart.\ Phys.\ } {\bf 12}, 107 (1999).

\bibitem{GENIUS}
H.~V.~Klapdor-Kleingrothaus,
arXiv:hep-ph/0104028.



\bibitem{neut}
G.~Jungman, M.~Kamionkowski and K.~Griest,
{\it Phys.\ Rept.\ } {\bf 267}, 195 (1996); \\
{\tt http://t8web.lanl.gov/people/jungman/neut-package.html}.


\end{thebibliography}
\end{document}